\documentclass[a4paper]{jpconf}
\usepackage{graphicx}
\begin{document}
\title{Loop quantum cosmology in the cosmic microwave background \\ \vspace*{0.3cm}
\small{International Conference on Gravitation and Cosmology, Goa, December 2011}}

\author{Julien Grain}

\address{Institut d'Astrophysique Spatiale, CNRS/Universit\'e Paris XI \\ Centre Universitaire d'Orsay, B\^at. 120-121, 91405 Orsay CEDEX, France}

\ead{julien.grain@ias.u-psud.fr}

\begin{abstract}
The primordial Universe can be used as a laboratory to set constraints on quantum gravity. In the framework of Loop Quantum Cosmology, we show that such a proposal for quantum gravity not only solves for the big bang singularity issu but also naturally generates inflation. Thanks to a quantitative computation of the amount of gravity waves produced in the loopy early Universe, we show that future cosmological datas on the polarized anisotropies of the Cosmic Microwave Background can be used to probe LQC model of the Universe.
\end{abstract}

\paragraph{Methodological introduction--} Building a quantum theory of gravity is probably the most outstanding problem of modern physics, but remains also one of the most difficult task. Facing such a great challenge --and retrospectively, looking at the many attempts developped since the last decades-- one could be tempted to consider that searching for quantum gravity is a never ending quest and that the wise attitude would be to withdraw such a challenge. However, quantum gravity is not {\it optional} ! Several heuristic arguments pin down the necessity of a successful building of quantum gravity \cite{oriti}. More importantly, the singularity theorems derived by Penrose and Hawking \cite{hawking} shows that general relativity ontologically fails: a way out reconciling quantum mechanics and general relativity is mandatory !

A new physical paradigm is needed and, to be successful, such a potentially new paradigm requires a well-established theoretical framework. On the one hand, this assumes a well-posed theoretical problem\footnote{By a well-posed theoretical problem we mean an {\it ill-posed} enigma with respect to establised theories.} --namely how to quantize general relativity or, alternatively, how to 'gravitize' quantum field theory? Many tentative theories have been proposed such as string theory \cite{string} and Loop Quantum Gravity (LQG) \cite{lqg} to mention the most popular ones, and though some of them have made great progresses, none of them is fully setted up. On the other hand, this also implies a new kind of paradigmatic experiments or observations. This is a mandatory step to discriminate theoretical proposals which are deeply on the wrong track from those which are not, as well as to guide theoretical developments. This well-posed phenomenological and experimental problem is the second main difficulty of quantum gravity as quantum gravitational effects should pop up at an energy scale of $M_\mathrm{Pl}\sim10^{19}$~GeV, far beyond any high energy physics experiments. Fortunately, the failure of general relativity precisely points towards those phenomena potentially probing quantum gravity: space-time singularities provide the territory for quantum gravity and one therefore has to search for those phenomena built up of singular space-times. This qualifies the primordial Universe, being the neighborhood of the big bang singularity, as an ideal territory for probing quantum gravity !

Observing quantum gravity is therefore splitted into two questions. How a given proposal for quantum gravity affects the physics of the primordial Universe {\it and} how can we probe the physics of the primordial Universe? Apparently, the second question points towards observation. But, as a theoretical question, this should not be overlooked since it strongly determined the precise implementation of the first question\footnote{As further discussed in the core of this note, this {\it observational} question requires the computation of cosmological perturbations. In the case of {\it e.g.} Loop Quantum Cosmology, implementing those perturbations on top of a quantum modified background space-time leads to theoretical developments to cancel anomalies in the algebra \cite{cailleteau}, and points toward a change of signature --that is an {\it euclideanization} of the background-- during the cosmic history \cite{bojopaily}.} The main requirement for a quantum gravity proposal is to solve for the pathological big bang singularities and this can be checked by computing the global evolution of the Universe in a quantum cosmological setting. However, it should be stressed out that such a global evolution cannot be probed directly as one cannot extract himself from the Universe to 'see' how it evolves in its primary ages. We are stuck inside the Universe and, from the inside, we are able to probe the physics of the early Universe via the Cosmic Microwave Background (CMB) anisotropies. The origin of such anisotropies are {\it cosmological perturbations} produced during a phase of accelerated expansion dubbed {\it cosmic inflation}. Inflation is a key ingredient of the standard cosmological paradigm as it solves for the horizon and flatness problems and provides a mechanism for generating perturbations, the primordial seeds for galaxies and large scale structures formation. Moreover, inflation could be a high energetic phenomenon (its energy scale could be as high as $\sim10^{16}$~GeV) and is therefore potentially affected by quantum gravity. Nevertheless, the inflationnary scenario is not free of any problems as it is very difficult to generate such a phase without invoking speculative physics or fine tuning.

Bridging a link between Loop Quantum Cosmology (LQC) --the cosmological implementation of LQG \cite{lqc}-- and inflation in the early Universe allows us to solve for the big bang singularity {\it and} to naturally trig a phase of inflation. In addition, a quantitative computation of cosmological perturbations produced in such a {\it loopy} universe makes the early Universe a potential laboratory for testing quantum gravity. Our methodology is the following. The evolution of the Universe is derived in the framework of {\it effective} LQC which includes first order quantum corrections coming from holonomies. On top of that modified cosmological background, we will quantitatively compute the amount of cosmological perturbations of tensor type ({\it i.e.} primordial gravity waves) and investigate how it impacts on the $B$-mode angular power spectrum of CMB anisotropies. Such an angular power spectrum is the main observable and can finally be used to forecast how future CMB observations could be used to probe LQC. We consider the early Universe to be filled with a massive scalar field $\Phi$ as matter sources and our time variable is cosmic time $t$.

\paragraph{Background evolution--} With holonomy corrections at first order, the modifed Friedman and Klein-Gordon equations are (denoting $H=\dot{a}/a$ the Hubble parameter and $m_\Phi$ the mass of the scalar field):
\begin{eqnarray}
	H^2=\frac{8\pi G}{3}\rho\left(1-\frac{\rho}{\rho_c}\right)&\mathrm{and}&\ddot{\Phi}+3H\dot\Phi+m^2_\Phi\Phi=0
\end{eqnarray}
where $\rho_c\sim0.8\times\rho_\mathrm{Pl}$ is a critical energy density which cannot be overcomed and which encodes LQC corrections. Because of  $(1-\rho/\rho_c)$, the Universe is not singular as the big bang is now replaced by a big bounce. The basical history of the Universe is thus a contracting phase, followed by an expanding phase. The {\it regular} transition from the contracting phase to the expanding one is ensured by the quantum corrections and the bounce occurs at $\rho=\rho_c$. More interestingly, this peculiar evolution of the Universe naturally leads to inflation ! Indeed, a phase of accelerated expansion can start right after the bounce if the scalar field is in the appropriate energy state for the slow-roll conditions to be fulfilled. For a massive field, this is translated into $\Phi(t_i)\sim 3.1~M_\mathrm{Pl}$ with $t_i$ denoting some time {\it after} the bounce. This condition is not easily met in the standard cosmological scenario. But in the LQC bouncing Universe, the Hubble parameter $H$ is negative valued during contraction. It therefore acts as an {\it anti-friction} term making $\Phi$ to climb up its potential. Right after the bounce, it appears that such a field is precisely in the appropriate energy setting for a sufficiently long phase of inflation to start \cite{jakub2010}. The key point about this scenario has been raised by Ashtekar and Sloan who showed that generating a phase of inflation in the LQC Universe with the mandatory amount of at least 60 e-folds is very close to unity, making inflation rather natural in this framework \cite{sloan}.

\paragraph{CMB power spectra--} To derive the amount of gravity waves produced at the end of inflation, one needs to solve the following equation describing tensor perturbations evolving on top of the modified FLRW background:
\begin{equation}
	\ddot{h}^i_a+3H\dot{h}^i_a-\frac{1}{a^2}\nabla^2h^i_a+12\pi G\frac{\rho}{\rho_c}\left(\frac{1}{3}\rho-V(\Phi)\right)h^i_a=0,
\end{equation}
where the last term, acting as a time-dependant mass term, encodes the 'quantum deformation' of the background. This equation of motion \cite{Bojowald:2007cd} has been derived from an algebra which is anomaly-free at all orders and 
can be safely used throughout the 
entire history of the bouncing universe\footnote{The issue of a closed algebra in effective LQC for both scalar and vector perturbations has been recently solved in \cite{cailleteau}.}. The main characteristics of a 'bouncy' power spectrum for tensor modes are the following. The IR (large scales) part is $k^2$ suppressed. This is due to the freezing of very large-scale modes in 
the Minkowski vacuum. Those modes indeed exit the horizon long {\it before} the bounce and
naturally exhibit a quadratic spectrum. The UV part is identical to the standard prediction. Small scales indeed 
experience a history basically similar to that of the big bang scenario. They exit the horizon
during inflation and reenter later, leading to the standard nearly scale-invariant spectrum. Intermediate scales, around $k\approx k_\star$, exhibit both a bump of 
amplitude $R$ and damped oscillations\footnote{The critical wavenumber (locating the bump and determining the typical scales for which the power spectrum changes from its $k^2$ shape to its nearly scale-invariant shape) is a phenomenological parameter related to $m_\Phi$ and to the fraction of potential energy at the time of the bounce. Similarly, $R$ is a phenomenological parameter related to $m_\Phi$ (see \cite{grain} for further details).}. This is mostly due to the fact that all modes are 
inevitably in causal contact at the bounce (the Hubble parameter vanishes, therefore leading
to an infinite Hubble radius). Those characteristics have been fully determined by numerically solving the equations of motion 
of tensor perturbations  propagating in the LQC-corrected, \{bouncing+inflationary\} 
universe \cite{jakub2010}. 
\begin{figure}
	\includegraphics[scale=0.65]{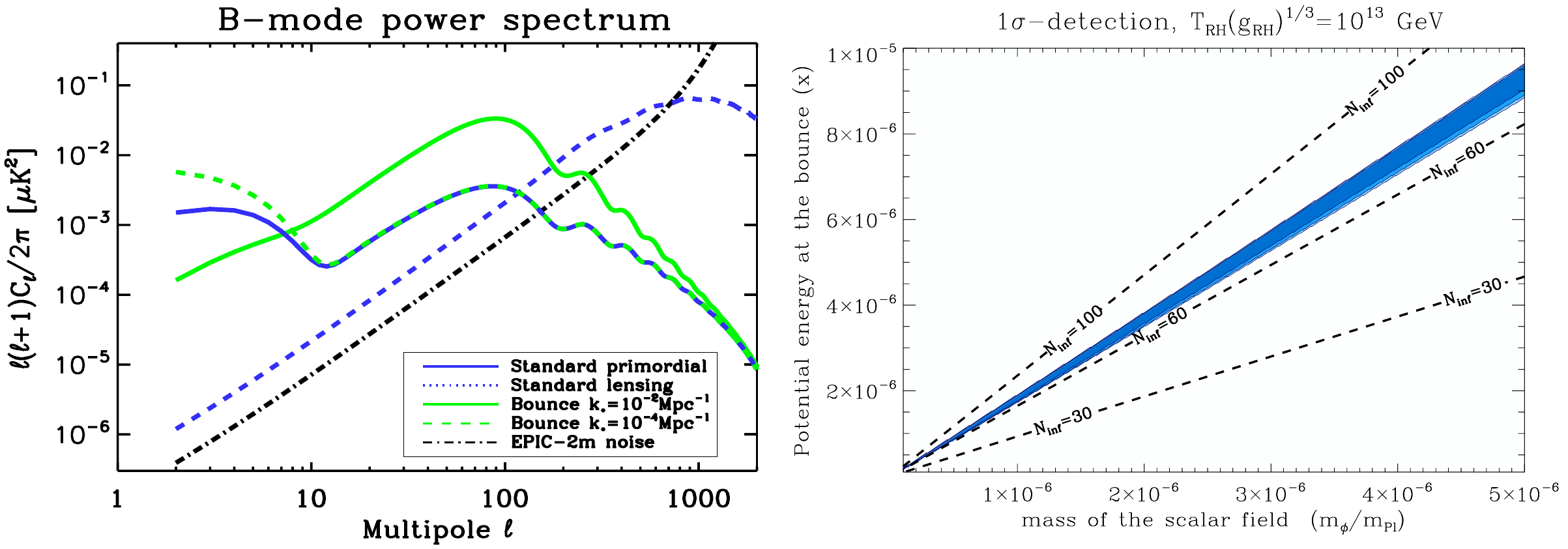}
	\caption{\label{fig} $B$-mode angular power spectrum (left panel) and detectable region of the parameters $(m_\Phi,x)$ describing a loopy universe (right panel). Figure adapted from \cite{grain}.}
\end{figure}

Using a Boltzmann code, such a primordial tensor power spectrum serves as an input for computing the footprints of primordial gravity waves on the CMB $B$-mode angular power spectrum (see left panel of Fig. \ref{fig}). As compared to the standard prediction, the distortion of LQC $B$-mode power spectrum depends on the value of $k_\star$ \cite{grain}. (Roughly speaking, the suppression of power at large scales can be 'seen' in the CMB only for values of $k_\star$ corresponding to scales {\it smaller} than the Hubble horizon today.) If $k_\star<k_\mathrm{Hubble}$ corresponding to a length scale greater than the Hubble scale today, the tail of the bump for $k>k_\star$ translates into a slight bump at large angular scales ({\it i.e.} small values of the multipole $\ell$). However, for $k_\star>k_\mathrm{Hubble}$, the $B$-mode angular spectrum shows, as compared to the standard prediction, a suppression at large angular scales due to the suppression at large length scales in the primordial spectrum and a bump at intermediate angular scales $\ell\sim100$. Finally, if $k_\star$ is greater than $\sim10^{-1}$ Mpc$^{-1}$, the suppression is effective up to $\ell$-values of a few hundreds and the primordial part of the $B$-mode is therefore systematically pushed below the lensing-induced $B$-mode.

Using those angular power spectra as an observational probe and considering both the instrumental noise as expected for a future dedicated $B$-mode experiment\footnote{The noise level is set to $2.2~\mu$K-arcmin., the beam width to 8 arcmin. and the sky fraction to 70\%.} and the lensing-induced $B$-mode as a 'foreground' masking the detection of primordial $B$-mode, one can forecast some constraints to be set on the fundamental parameters describing the LQC model of the Universe \cite{grain}. The detectable values at 1-$\sigma$ of the mass of the scalar field $m_\Phi$ and its fraction of potential energy at the bounce, denoted $x$, are depicted by the blue band in the right panel of Fig. \ref{fig}. This roughly corresponds to a detectable range of $k_\star$ from $2\times10^{-4}$~Mpc$^{-1}$ to $3\times10^{-1}$~Mpc$^{-1}$ considering {\it degeneracies} with other cosmological parameters. The upper part is not detectable as it corresponds to $k_\star\ll k_\mathrm{Hubble}$ making the $B$-mode power spectrum {\it undistorted} as compared to the standard general relativistic prediction. The lower part is not detectable as it corresponds to $k_\star\gg10^{-1}$ making the primordial $B$-mode systematically smaller than the lensing-induced part. Though measurements of the LQC parameters is not possible in this second case, a discrimination with pure general relativity is still possible as the suppression induced by the bounce is 'seen' via the masking of the primordial $B$-mode\footnote{This 'suppression' could also be interpreted as a very low value of the tensor-to-scalar ratio $T/S$. However in the precise case of LQC, the three parameters $T/S$, $k_\star$ and $R$ are fully determined by the {\it two} parameters $m_\Phi$ and $x$, thus breaking the degeneracy between $k_\star$ and $T/S$ (see \cite{grain} for details).}; that is via its {\it non-detection}.

\paragraph{Conclusion--} LQG offers an appealing approach to build a quantum theory of gravity. Its implementation to the symmetry reduced case of FLRW metric describing our whole Universe shows that the big bang singularity is cured and that inflation is {\it naturally} triggered. Moreover, thanks to a quantitative computation of gravity waves produced in the early Universe, subsequently impacting on the CMB polarized anisotropies of $B$-type, a possible window to constraint those loopy models of the Universe using cosmological datas is now opened. As stressed out in \cite{essay}, the case of LQC illustrates a possible future way to test for quantum gravity using cosmological/astronomical observations.

\ack
The author warmly thanks his collaborators Aur\'elien Barrau, Thomas Cailleteau and Jakub Mielczarek.

\end{document}